\documentclass[%
 reprint,
 amsmath,amssymb,
 aps,
]{revtex4-2}
\usepackage{graphicx}
\usepackage{dcolumn}
\usepackage{bm}
\usepackage{color} 
\usepackage{cancel}
\usepackage{array}
\usepackage{tabularx}
\usepackage{booktabs}
\usepackage{graphicx}
\usepackage{dcolumn}
\usepackage{braket}
\makeatletter
\newcommand{\tpmod}[1]{{\@displayfalse\pmod{#1}}}
\makeatother

\begin{document}

\title{The role of Coulomb interaction on the electronic 
properties of monolayer NiX$_{2}$ \\ (X = S, Se):
A DFT+U+V study}

\author{Sergio Bravo}
\email{sergio.bravoc@usm.cl}
\address{Departamento de F\' isica, Universidad T\'ecnica
Federico Santa Mar\' ia, Av. Espana 1680, Casilla Postal
110V,Valpara\' iso, Chile}

\author{P.A. Orellana}
\address{Departamento de F\' isica, Universidad T\'ecnica
Federico Santa Mar\' ia, Av. Espana 1680, Casilla Postal
110V,Valpara\' iso, Chile}

\author{L. Rosales}
\address{Departamento de F\' isica, Universidad T\'ecnica
Federico Santa Mar\' ia, Av. Espana 1680, Casilla Postal
110V,Valpara\' iso, Chile}


\date{\today}

\begin{abstract}

The electronic structure of Nickel dichalcogenides, NiS$_2$
and NiSe$_2$, in monolayer form, is studied employing
first-principles methods.
We assess the importance of band ordering, covalency and
Coulomb interactions in the ground state of these systems. 
Hybrid functional results are compared with standard 
functionals and also with Hubbard-corrected functionals
to systematically address the role of electronic 
interactions and localization. We found
that mean-field correlation realized by intersite Hubbard 
interactions are directly linked to 
the magnitude of the energy band gap, giving
compelling evidence for the presence of a charge transfer
insulating phase in these materials.

\end{abstract}


\maketitle

\section{Introduction \protect\\} 

Two-dimensional (2D) Van der Waals materials have emerged
as a class of systems that have proven to hold great
potential for realizing new physics as well as novel
technological capabilities. One of the sets that has
aroused a high level of interest in recent times are
the transition metal dichalcogenides (TMD). They exhibit
an attractive combination of atomic-scale thickness, 
strong spin-orbit coupling, direct bandgap, and
favorable electronic and mechanical properties
\cite{2017NatRM...217033M}.
This makes them interesting for basic research and 
for example, high-end electronics and spintronics
applications.

Even with these advances, there remain many 2D TMD 
that deserve to be investigated in more detail.
To this category belong the Nickel dichalcogenides
NiX$_2$, with X=S, Se, Te. This group of materials has
been proposed theoretically \cite{2DMATPed,2dmatDB}
in the 2H and 1T phases and also realized experimentally,
in the case of 1T NiSe$_2$ \cite{NiSe2_exp}.
The properties studied are related to band engineering
\cite{BEng_NiS2}, 
thermoelectric efficiency \cite{1T-NiS2_TE}, 
anode materials in batteries and 
also superconductivity \cite{NiX2_anode}.
One of the ingredients that is lacking in these recent
studies is the inclusion of electronic interactions.
As we know, the presence of localized $d$ orbitals in 
transition metals make necessary that the 
correlation effects are accounted for so that the 
predicted physical properties are closer to reality.
The level of refinement in the treatment of interactions
varies widely, and factors that impact the selection of 
a method stems from the size of the system, the type of the 
components and the available computational resources.
Examples of methods that treat the interacting problem
are the Hartree-Fock approximation \cite{Levine-QC}, 
GW approximation \cite{IntElec}, 
mean-field approximations \cite{CMFT_book}, 
density functional theory \cite{DFT-martin}, 
dynamical mean-field theory \cite{IntElec}
and coupled cluster theory \cite{CClustTh}, which form
just a limited list.
Among the possible methods mentioned, mean-field methods
along with density functional theory (DFT), are the fastest
and most flexible in computational terms. This allows to
study systems with considerable sizes and different 
kinds of structure without high cost. 
Inside the DFT framework, one of the implementations of
electronic interactions at the mean-field level that is 
widely used, is known as Hubbard-corrected DFT, or
DFT+U, for its early implementation 
\cite{DFT+U_Anis,DFT+U-Coco}.
This formalism adds a minimal term to the original 
DFT energy functional which accounts for 
onsite Coulomb repulsion (U) at the atoms with localized
orbitals\cite{DFT+U-Coco}. This approach has been very
successful, and in recent years, it has been implemented
in extended form by including not only onsite parameters
but also intersite interactions 
\cite{DFTUV_Campo,DFT-UV-v2,DFT-UV-v3}.
This last method is starting to be used in studies of
different materials
\cite{DFTUV_2019,DFTUV-2020,DFTUV-2021,DFTUV-2022_1,
DFTUV-2022_2,DFTUV_2023},
giving new perspectives on the incidence of 
interactions on the electronic structure. 

To study the effects of correlation in
NiX$_2$ systems, in this work, we employ DFT with
Hubbard corrections and also with hybrid functionals
in order to characterize NiS$_2$ and NiSe$_2$ in 
the 1T monolayer form. We carry out a band structure
study in momentum space along with real space orbital 
characterization to explore the role of Coulomb
interaction in these systems at the mean-field level. 
Complementary, a charge transfer phase is identified
in both materials, and its origin is discussed.

The manuscript is organized as follows. 
First, we detail the computational resources 
and corresponding settings for the 
first-principles calculations. Next, we mention 
the more salient structural properties of 
NiX$_2$. The calculations using the 
non-corrected DFT methods are presented in section
\ref{DFT0} along with the results of 
the electronic structure using a hybrid functional.
Also, we present a discussion of the basic electronic
features. 
This is followed in \ref{LRcalc} by the main
calculation of the work. Namely, the application of
the Hubbard-corrected DFT method, which is
accompanied by the self-consistent calculation of
the Hubbard parameters by a linear response
approach. Comparison of this last procedure 
with the hybrid functional outcome is addressed
and in section \ref{LR-newV} a possible route to 
improve the Hubbard-corrected results is 
developed. We conclude the article by exposing final
remarks and giving an outlook for future work. 
Additional data and figures that complement the 
results and discussion presented in the main 
text has been left as supplementary material (SM).

\begin{figure*}
\includegraphics[width=1.5\columnwidth,clip]{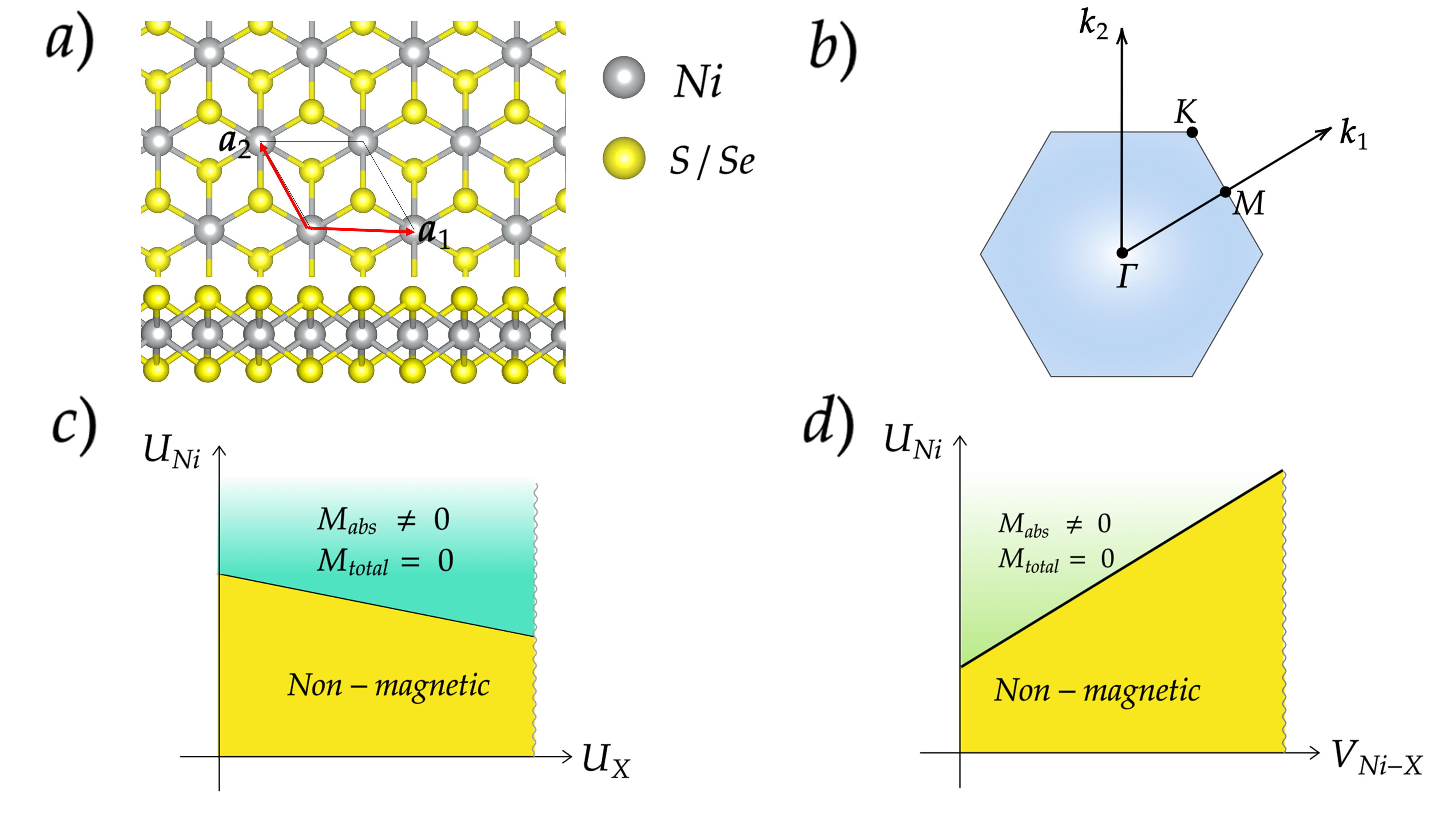}
\caption{a) The lattice structure of monolayer NiX$_2$.
b) Two-dimensional Brillouin zone for the monolayer
NiX$_2$ systems. 
c) Magnetic phase diagram for $U_{Ni}$ versus $U_S$ for
fixed intersite interaction $V_{ij}$. The range of values for 
$U_{Ni}$ was from 0 eV to 9 eV. For $U_X$ was from 0 to 4 eV.
d) Magnetic phase diagram for $U_{Ni}$ versus $V_{NiX}$ for fixed
onsite interaction $U_X$. The range of values for $U_{Ni}$ was
from 0 eV to 9 eV and $V_{NiX}$ was from 0 to 3 eV.
 }
\label{FIG-1}
\end{figure*}

\section{Computational details}\label{II}

All calculations were carried out using the QUANTUM ESPRESSO
package (QE) \cite{QE_2017}. 
We use three types of functionals. Namely, the standard GGA-PBE
functional, a modified PBE functional with Hubbard corrections
(also called DFT+U+V) according to implementations
in \cite{DFTUV_Campo} and the hybrid HSE06 (HSE) functional
\cite{HSE} with the ACE implementation \cite{ACE_hyb}. 
Relaxed structures were obtained at PBE and DFT+U+V levels
with a force tolerance of $10^{-2}$ eV/{\AA} per atom and
an energy tolerance of $10^{-8}$ Ry.
A vacuum layer of 20 {\AA} was used to simulate the
layered character of the systems. The energy cutoff for
the self-consistent calculations was set to 100 Ry for
all cases with a tolerance of $10^{-8}$ Ry. 
Calculations included the possibility of magnetic final states
without considering spin-orbit coupling, that is, a 
spin-polarized setting.
The $k$-space Monkhorst-Pack grid was fixed to
15 $\times$ 15 $\times$ 1 points for all calculations. 
In the case of HSE functional, the Fock operator was calculated
with a q-space grid of 5 $\times$ 5 $\times$ 1 points.

For the linear response calculations that allow the 
self-consistent computation of the Hubbard parameters, the
HP code was used \cite{HP_code}, as part of the QE suite. 
The code requires a self-consistent ground state calculation of
QE as starting point. 
We use the ortho-atomic type of projectors for this calculation,
which sets the basis for the occupations matrices
(see section \ref{LRcalc}) with L\"owdin orthogonalized
atomic orbitals. 
Well-converged results for the values of the parameters were
obtained using a q-space grid with 5 $\times$ 5 $\times$ 1 
points and a tolerance of $10^{-7}$ eV for the response function
$\chi$, defined in \cite{HP_code}.

\section{Results}

\subsection{Structural properties}

Nickel chalcogenides NiX$_2$ in 1T phase can crystallize in 
monolayer form with a trigonal lattice structure, as shown
in Fig. \ref{FIG-1}a.
The transition metal is located at the corners of the unit
cell forming a triangular sub-lattice while the chalcogen atoms
form a buckled hexagonal sub-lattice. This structure is 
described by space group (SG) $P\bar{3}m1$ (\#164) 
\cite{Findsymm}. 
This symmorphic SG has 12 symmetries generated by spatial
inversion $I$, a three-fold rotation about an axis perpendicular
to the monolayer plane, and a two-fold rotation about an axis
that is in the monolayer plane along the lattice vector
$\boldsymbol{a_1} + \boldsymbol{a_2}$ (see Fig. \ref{FIG-1}a 
for the lattice vectors representation).
The relaxed structure obtained from our calculations gives that
a generic Ni atom is six-fold coordinated, and the NiX$_6$
subsystem forms a distorted octahedron. Thus, for instance, using 
NiSe$_2$, we obtained that the elongated Se-Ni-Se angle has a
magnitude of 95.58°, departing from the ideal right angle.
Additional data concerning lattice parameters and bond lengths
arising from the first-principles calculations detailed below, 
are presented in the SM. The values reported in this work
agree well with previous studies
\cite{2dmatDB,NiX2_anode,NiSe2_exp}.

One important consequence of the atomic landscape is that Ni
atoms experience the well-known crystal field splitting associated
to the interaction with chalcogen orbitals. This splitting can
be directly deduced from the SG information by taking into
account that the atom located at the unit cell corner
(1a Wyckoff position to be more precise \cite{bradleyGT}), have
a site symmetry isomorphic to point group (PG) $\bar{3}m1$.
The dimensionality of the irreducible representations of this
PG gives the maximal degeneracy that a set of orbitals can have. 
Using the tabulated information from the Bilbao 
Crystallographic Server \cite{BCS_1} we can see
that the greatest degeneracy that orbitals can have in this
the atomic environment is two.
If we take into account the basis of atomic orbitals, we can
see that, for example, $d$-orbitals will split into
three sets; one nondegenerate orbital and two sets
of orbitals each with a double degeneracy.
The presence of these two-fold degenerated orbitals
will induce two-fold degeneracies in momentum space.
This real space momentum space relation will dictate
the general form that the electronic structure will show in 
what follows. 

\subsection{Orbital order and first electronic structure
description}
\label{DFT0}

\begin{figure*}
\includegraphics[width=1.6\columnwidth,clip]{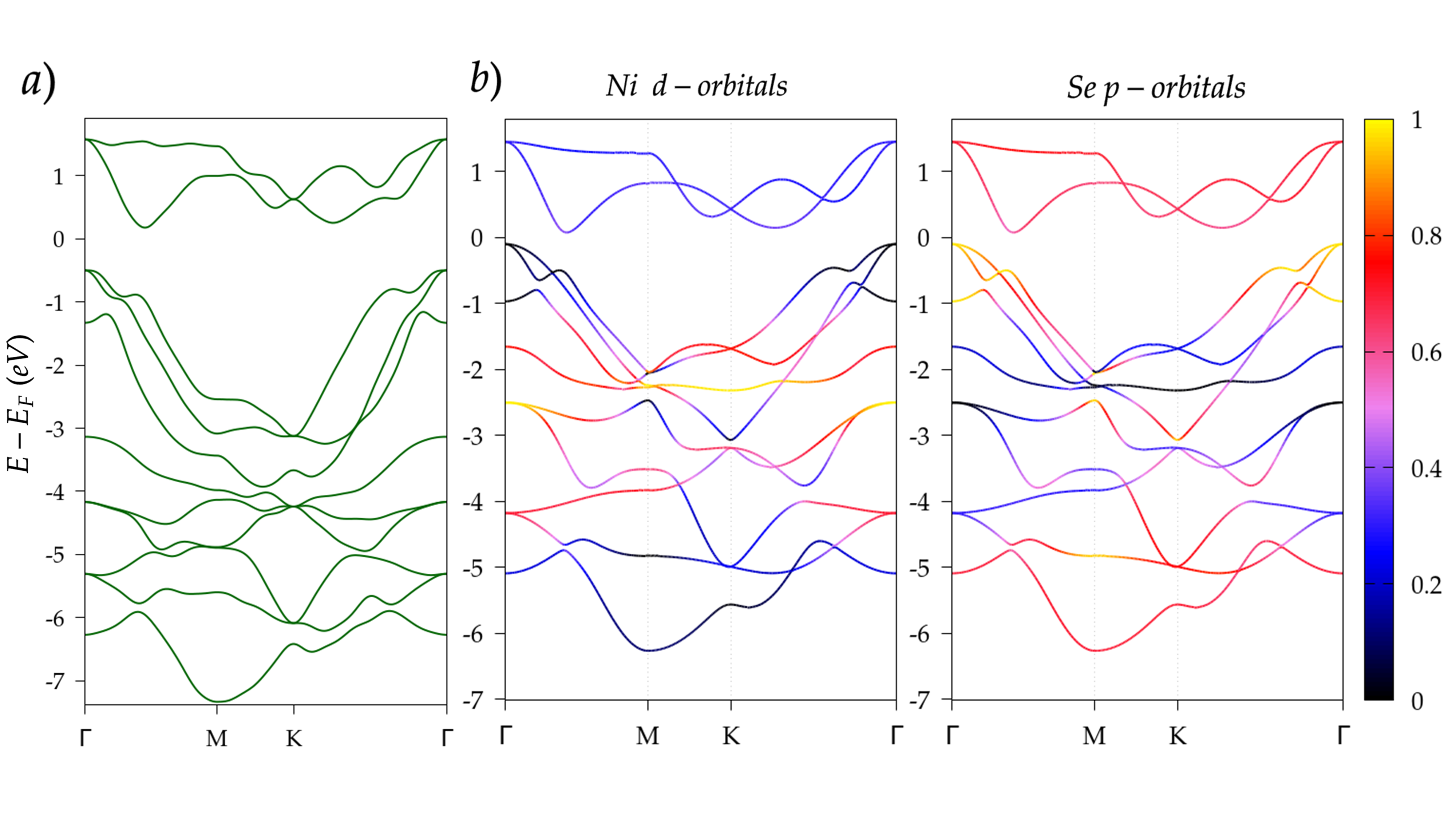}
\caption{a) HSE electronic band structure for NiSe$_2$. 
b) PBE electronic band structure with orbital projections for
(left) Ni $d$-orbitals and (right) Se $p$-orbitals.
The color scale at the right is in states $\cdot$ A$^2$/eV units.}
\label{FIG-2}
\end{figure*}

\begin{figure}[ht]
\includegraphics[width=1.\columnwidth,clip]{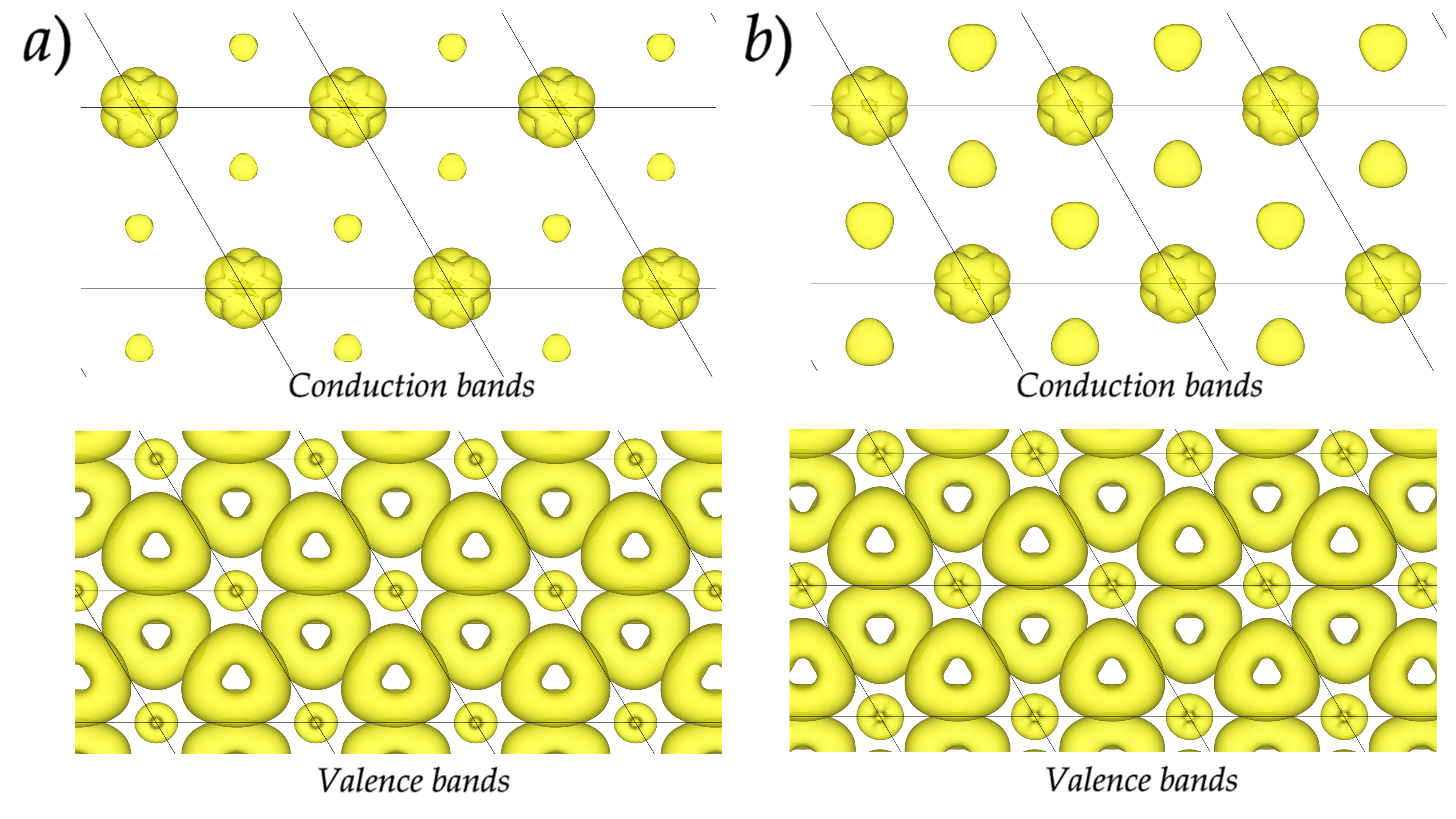}
\caption{a) Integrated local density of states for (top panel)
lowest conduction bands and (bottom panel) top of the valence 
bands with the HSE functional for NiSe$_2$. 
b) Integrated local density of states for (top panel)
lowest conduction bands and (bottom panel) top of the valence 
bands with the PBE functional for NiSe$_2$. In both cases, the 
valence band range is 0.4 eV from the top of the bands. The
conduction band integration range goes from Fermi level
to 1.5 eV.}
\label{FIG-3}
\end{figure}

Although correlated systems could be studied by more elaborated 
methods \cite{IntElec}, we take here a mean-field approach, as
this can serve as a baseline for understanding the incidence
of interactions in these TMD. 
The standard starting point of this approach is to perform an 
electronic structure at the level of the generalized gradient 
approximation (GGA) with the PBE functional. In the following, 
we present results for the NiSe$_2$ monolayer and refer the reader
to the SM for analogous figures and data concerning the 
NiS$_2$ monolayer.
In first place, we present the NiSe$_2$ electronic band structure
in Fig. \ref{FIG-2}b.
The plot also contains information about the orbital projection
along the high-symmetry path delimited by the high-symmetry 
points (HSP) in the Brillouin zone (BZ), as represented in
Fig. \ref{FIG-1}b. 
The PBE projected bands show a clear tendency even at this 
level; upper valence bands show a clear $p$-orbital type
coming from Se atoms. 
However, deep valence bands have a more mixed character,
they are formed by the strong hybridization of Ni $d$-orbitals
and Se $p$-orbitals, as expected by the features of the
lattice structure.
This orbital energy order is similar to the behavior observed
in charge-transfer insulators \cite{khomskii-book}. 
Specifically, when the $p$ orbitals are on top of the
$d$ orbitals in the valence bands, the material will
present a small o even negative charge-transfer
gap \cite{khomskii-book}.
This charge transfer can be understood as the passing of
electronic charge from the $p$-shell at chalcogen atoms
to the $d$-shell belonging to the transition metal. This
is formally represented as 
$d^{n}p^6 \rightarrow d^{n+1}p^5$ and in this type
of systems, this process will correspond to the
low-energy excitations \cite{Nio_PBR2020}.
However, as it is well-known, PBE results consistently 
underestimates localization effects and systematically tends
to delocalize electrons \cite{DFT-martin} which results in
a sometimes deficient prediction of the ground state properties.
For transition metal compounds, this is a well-documented 
issue and methods to improve the PBE results are needed
to meet experiments \cite{DFT+U-Coco}.
Following this line, we will explore two levels of extension
of the PBE results. 
In this section, we present the first one. 
In Fig. \ref{FIG-2}a, the electronic band structure resulting
from using a hybrid HSE functional is shown. 
This functional, by definition, includes a part of the exact
exchange \cite{HSE}. 
The hybrid band structure exhibits a gap of approximately 1 eV, 
which, as expected, is greater than the PBE band gap, which
has a magnitude of $\sim$ 0.6 eV.
Another difference stems from the dispersion of top valence
bands and their interaction with deep valence bands. 
It can be observed that the HSE result yields a more clear
separation between the set of upper and lower valence bands
and higher dispersion for the upper valence bands.
These effects can be traced to including 
localization in the hybrid case. This can be observed if we study 
the real space orbital character of NiSe$_2$, using the local 
density of states. More in detail, in Fig \ref{FIG-3} we compare
the level of localization for the low energy range of the system 
obtained from the hybrid functional versus the PBE result,
represented by the integrated local density of states (ILDOS). 
Although the results coincide roughly, a closer look show that the
$p$-orbitals contribution to the conduction bands is more localized
in the HSE case (Fig \ref{FIG-3}a top) with respect to the PBE
result (Fig. \ref{FIG-3}b top). Also, for the upper valence bands, 
states situated at Ni atoms show a slightly more delocalized
appearance in PBE (Fig. \ref{FIG-3}b lower panel) in
comparison with the HSE case (Fig. \ref{FIG-3}a, lower panel).
All the above discussion indicates that the inclusion of electron
correlations in monolayer NiX$_2$ materials should play a key role
to describe properties derived from the electronic structure.
In principle, we can stay with the hybrid functional as the final
result. However, hybrid functionals suffer from technical 
complications related to the computational effort to calculate
the exact exchange and this fact hinders their utilization
in wider applications. Also this kind of functionals also
lack other contributions to correlation \cite{IntElec} and
thus more sophisticated approaches should be used, such as 
GW calculations and dynamical mean-field methods.  
Despite of this, the HSE result represents a significant 
improvement over the bare PBE calculation. 
Thus, in the following, we take the HSE ground state as a 
reference and compare it with another method that tries 
to improve on the PBE level by including Coulomb interactions
at the mean-field level. 

\subsection{DFT+U+V approach and linear response calculations}
\label{LRcalc}

We have calculated the electronic structure of NiX$_2$ materials
considering the inclusion of correlation through Hubbard 
parameters. 
We introduce the parameters by resorting to a simplified model
to visualize this setting better.
From the above calculations, we can identify the 
Ni $d$-orbitals and the Se (S) $p$-orbitals as the most
important orbitals.
Using a tight-binding representation with no correlation, the 
low energy Hamiltonian takes the following form

\begin{equation}
\begin{aligned}
H_0=\epsilon_d & \sum_{\alpha} d_{\alpha}^{\dagger} d_{\alpha}
+\epsilon_p \sum_{\beta} p_{\beta}^{\dagger} p_{\beta} \\
& +\sum_{\alpha \beta} t_{\alpha \beta}\left(d_{\alpha}^{\dagger}p_{\beta}
+\text{ h.c. }\right).
\end{aligned}
\end{equation}

Here $d_{\alpha}^{\dagger}$ and $p_{\beta}^{\dagger}$ are
the creation operators for an electron in $d$-orbitals
and $p$-orbitals, respectively. 
The terms in the first line correspond to onsite
energies and the last term is the hopping parameter that serves
to implement hybridization among Ni and Se orbitals.
The inclusion of correlation is realized as an extended 
Hubbard model that is expressed as \cite{khomskii-book}

\begin{equation}
\begin{aligned}
H_{UV} =& H_0 + U_{Ni} \sum_{\alpha} n_{d_\alpha}n_{d_\alpha}
+ U_{X} \sum n_{p_\beta} n_{p_\beta}\\
&+V_{NiX}\sum_{<\alpha\beta>} n_{d_\alpha}n_{p_\beta} 
+ V_{XX}\sum_{<<\beta\gamma>>} n_{p_\beta}n_{p_\gamma}, 
\end{aligned}
\end{equation}
where $X = S,Se$.

In this Hamiltonian, $n_{d_\alpha}$ and $n_{p_\beta}$ denote
the occupation operators for $d$ and $p$ orbitals, respectively. 
The second and third terms represent the so-called onsite
Hubbard parameters that quantify the cost in energy of
double occupancy for a $d$-orbital ($U_{Ni}$) and a 
$p$-orbital ($U_X$). The fourth term, with coefficient
$V_{NiX}$, represents an extension of the theory to include
the intersite interaction between occupations among the
$d$ and $p$-orbitals. 
The last term is also an intersite interaction between
the next-nearest neighbors of the Se-Se type in this
class of materials. In practice, we include two of these
(Se-Se) interactions; second and third neighbors. We stop
at this level and leave more distant correlations
outside the subsequent calculations. 

A density functional theory implementation of the 
interactions presented above, with a self-consistent 
and rotational invariant character is employed for the 
calculation of the ground states. This level of theory 
is customarily denoted as DFT+U+V \cite{DFTUV_Campo}. 
We do not enter into the details of the theory here 
as this has been extensively exposed in recent works 
\cite{DFTUV_2019,DFTUV-2020}. We only mention that the
formalism is based on the modification of the original
DFT framework by the addition of a correction term
$E_{U,V}$ giving a new energy functional
$E = E_DFT + E_{U,v}$.
The key ingredients of the $E_{U,V}$ term are the
so-called occupation matrices
$n_{m, m^{\prime}}^{IJ\sigma}$ which are defined
by the projection of the Kohn-Sham eigenfunctions
$\psi_{\mu}^\sigma$ ($\sigma$ labels spin ) on a set
of localized orbitals $\phi_m^I$ ($J$ labels
the atomic site and $m$ labels the orbital quantum number),
such that \cite{DFT+U-Coco}

\begin{equation}
n_{m, m^{\prime}}^{IJ\sigma}=\sum_{\mu} f_{\mu}^\sigma
\left\langle\psi_{\mu}^\sigma \mid
\phi_{m^{\prime}}^J\right\rangle
\left\langle\phi_m^I \mid \psi_{\mu}^\sigma\right\rangle.
\end{equation}

The coefficients accompanying these matrices are the Hubbard
onsite and intersite parameters presented above. 
The $n_{m, m^{\prime}}^{IJ\sigma}$ are obtained in a 
self-consistently way along the energy computation and
comprise an appropriate model for interactions at the
mean-field level, with the advantage of being part of a fast
and flexible framework.

As has been made patent by the model presented, the Hubbard
parameters will modify the localization character of the 
orbital manifolds of interest. However, until recently, 
the most used procedure was to set the parameters in a 
semi-empirical way, resorting to experimental references
or previous numerical benchmarks \cite{U-bayesian}. 
Presently, the approach is to calculate these
parameters in a self-consistent
manner and various alternatives have been made available 
\cite{DFTU_PRL,DFT-UV-v2,DFT-UV-v3}.

For this work, we use the DFT+U+V approach as formulated in 
\cite{LR-Timrov}.
The Hubbard parameters are computed within this framework using
density functional perturbation theory, also known as a linear
response (LR) calculation. Section \ref{II} presents the specific
settings used for the calculations in this work.
We now outline the procedure that has been put forward for
procuring a converged set of parameters. In the first
place, the calculation needs an initial guess for 
the Hubbard parameters. 
We have started from zero values and, with this set, run
a first calculation.
One of the subtle points is that our starting configuration
concerns an insulating phase with no magnetic state. 
However, as it is well-known, variations of the electron
interaction strength could result in magnetic states
and also in metal-insulator transitions \cite{khomskii-book}.
Thereby, after the initial LR computation (first shot), we
carried a new ground state computation with the obtained 
set of parameters. 
This calculation shows that the actual
ground state is metallic and magnetic at this point of the 
process. The material also suffers from an
enlargement of the unit cell due to the 
increment of the interactions
(See the SM for a summary of the geometric parameters obtained 
for the magnetic phases). 
We therefore, go to obtain new relaxed coordinates. 
This ground state is now used as the input of a second shot
for the LR calculation to see how parameters vary. 
After this shot, the procedure explained after the first shot
is repeated until the magnitude of the parameters does not
change above a desired tolerance. 
For our purposes, we set the threshold
at 0.1 eV, where we noted that the electronic structure does
not suffer appreciable changes for all the considered
parameters. This iterative process yields the
LR Hubbard parameters that are presented in 
TABLE \ref{LR-tab}. 
The final phase obtained from LR is a magnetic metallic 
state. In particular, we find a compensated ferrimagnetic
(FiM) ground state, which has a zero total 
magnetization and a nonzero absolute magnetization
of $\sim 2.7$ $\mu_B/$cell for NiS$_2$ and 
$\sim 3.1$ $\mu_B/$cell for NiSe$_2$.
The microscopic magnetic ordering that is realized entails
an antiferromagnetic (AFM) interaction among the Ni atoms
and the chalcogen atoms. 
This AFM coupling is complemented with a complete
compensation of magnetic moments in the unit cell, 
resulting in zero total magnetization.
The electronic band structure that is obtained for 
this phase is presented in the SM.   

In view that a double transition has occurred in the system,
we decided to carry out a study of the magnetic
properties of the system. We achieve this through 
performing several DFT+U+V calculations within a range of 
values for U and V parameters (See caption of Fig. 
\ref{FIG-1}a and b). 
To achieve this, we conduct multiple DFT+U+V calculations
using different values for the U and V parameters
(as indicated in the caption of Fig. \ref{FIG-1}a and b).
Observation of the trends in these calculations indicates
that the FiM state is very robust, as it is the only 
ground state encountered for the range of values explored 
(without considering eventual lattice distortions that
could change the space group symmetry). 
This FiM state appears even in the case where only 
$U_{Ni}$ is nonzero. 
Considering the former and the tendency to an AFM
coupling, we can expect that the final FiM state will
be produced due to the difference in localization of 
the $p$ electrons in comparison with $d$ electrons; the
more itinerant $p$ electrons tend to metalize the system,
while $d$ electrons will tend to stay localized. In the 
overall, for a sufficiently high $U_{Ni}$ interaction
(of the order of $\sim 6.5$ eV for both
materials), it will be favorable to produce a conducting
channel with only one type of spin polarization, leaving
the remaining channel insulating.
Therefore in comparison with our reference calculation, 
the hybrid functional result of section \ref{DFT0}, 
the LR is not giving the correct ground state.
Nevertheless, the inspection of the magnetic phase diagrams
points to the fact that the state similar to the
reference calculation must be located in the nonmagnetic
zone but with nonzero interaction parameters. 
If the resulting values of the LR calculation are
considered and compared with values extensively used
\cite{U-Values_PRB92,U-values_PRB},
it can be concluded that the actual magnitude of the
onsite Hubbard parameters, $U_{Ni}$ and $U_{X}$ are
in good agreement with previous works. This is not 
the case for the intersite parameters, due to the
scarce availability of results. 
Thus in the next, we will explore the implications
of keeping onsite values fixed while varying the
intersite parameters.
A guideline that this is indeed plausible is that, 
observing the $U_{Ni}$ versus $V_{NiX}$ diagram in
Fig. \ref{FIG-1}d, if we start in the FiM region
and move horizontally in the graph to greater
values of $V_{NiX}$, we end invariably in a
nonmagnetic correlated state.

\begin{table}[]
\resizebox{\columnwidth}{!}{%
\begin{tabular}{ccccccc}
\hline
Material & $U_{Ni}$ & $U_{X}$ & $V_{NiX}$ & $V_{XX}^{(1)}$ &
$V_{XX}^{(2)}$ & $\bar{V}_{NiX}$ \\ \hline
NiS$_2$  & 6.79     & 3.84    & 0.68      & 0.51      & 0.47
& 1.8            \\
NiSe$_2$ & 6.5     & 3.53    & 0.61      & 0.45      & 0.41
& 2.3            
\end{tabular}%
}
\caption{Hubbard parameters obtained from linear response
calculations for monolayer NiS$_2$ and NiSe$_2$. The last
column reports the modified values used for the first
neighbor intersite interactions. All values are in eV units.}
\label{LR-tab}
\end{table}

\begin{figure*}[ht]
\includegraphics[width=1.7\columnwidth,clip]{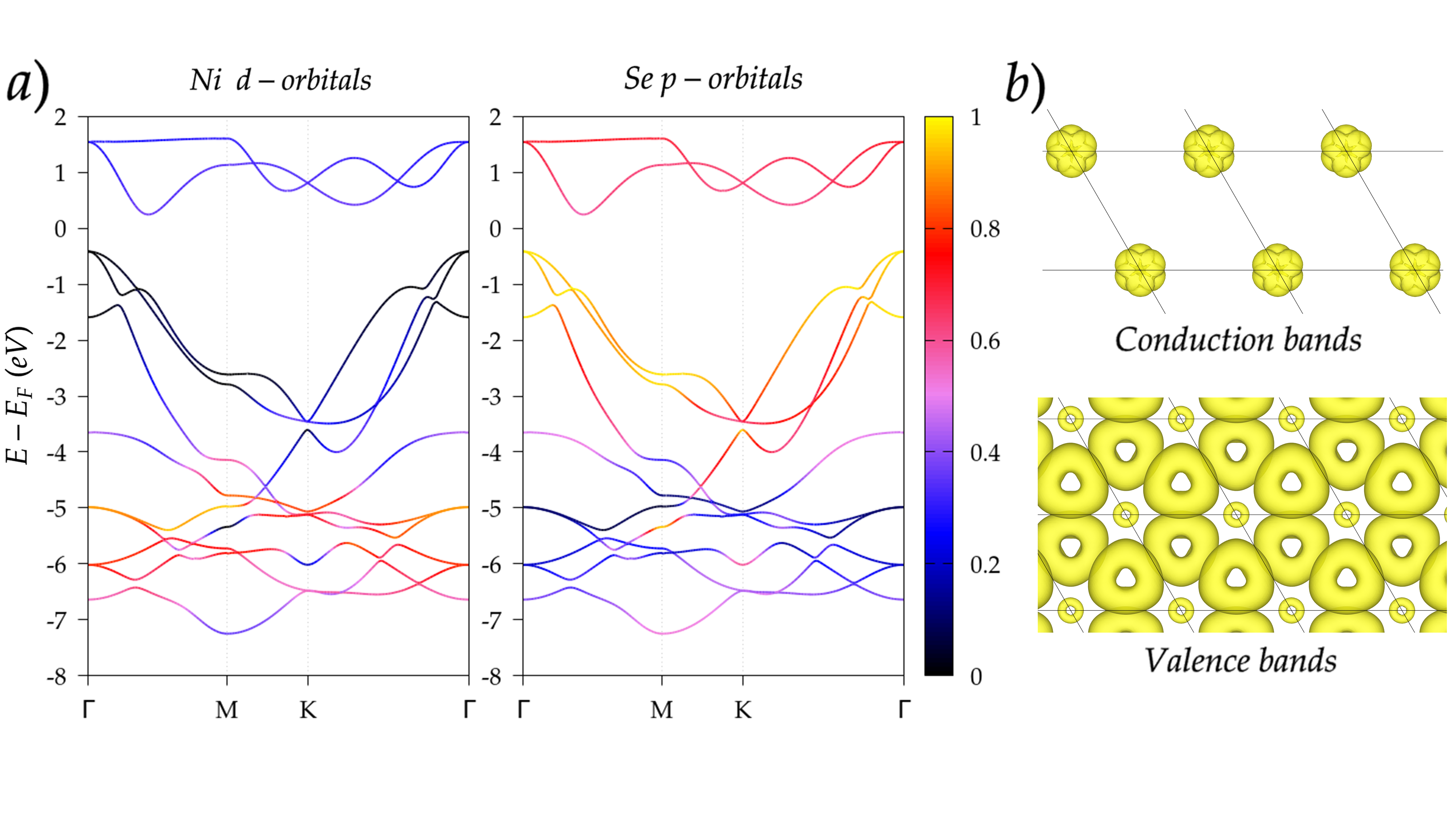}
\caption{a) Orbital-projected electronic band structure at the
DFT+U+V level of NiSe$_2$ with parameters from linear response 
with modified $V_{NiX}$ parameter. 
The color scale at the right is in (states $\cdot$ \AA$^2$)/eV units.
b) Integrated local density of states for (top panel)
lowest conduction bands and (bottom panel) top of the valence
bands using the DFT+U+V functional with linear response parameters
with modified $V_{NiX}$.
}
\label{FIG-4}
\end{figure*}

\subsection{Control of the energy gap by minimal modification of 
the linear response result}\label{LR-newV}

As mentioned in the previous section, we will examine how 
to modify the results from LR to attain a more similar
state to that of our reference hybrid functional. 
The idea, the basis of which was established above, is only
to alter the inter-site parameters. This is in the spirit of 
minimizing the arbitrary adjustments in the study.
To constrain even more the approach, we mention that,
from the calculations mentioned in the previous section, 
the $V_{XX}$ interactions do not have a substantial impact
on magnetic properties of the systems. Therefore, we propose a 
a minimal modification that will consist of to only change
the $V_{NiX}$ parameter, the other parameters remaining
at the same values as in the original LR calculation.
We have increased $V_{NiX}$ beyond values where a 
magnetic phase transition is produced, implying a 
metal-insulator transition.
Both transitions happen at almost the same $V_{NiX}$ values.
The trend observed in the process is that the absolute 
magnetization decreases when $V_{NiX}$ grows. This happens
up to a point where it abruptly goes to zero. The critical
value found was $\sim 1.6 eV$ for NiS$_2$ and $\sim 1.8 eV$
for NiSe$_2$.
This magnetic transition is also accompanied by a reduction
of the lattice parameter of the unit cell in such a way that, 
after the transition, the system recovers the original 
lattice constant found in the non-interacting 
GGA calculation. 

As a result, the system will no longer have magnetic properties
and will display an energy gap beyond this point. At this
stage, the most notable aspect is that as the parameter
increases, there is an increase in the energy gap of the
system. In simpler terms, there is a direct relationship
between the value of $V_{NiX}$ and the size of the gap.
To compare with the hybrid case, our criterion for setting
the value of $V_{NiX}$ is to match the energy gap magnitude
obtained from the HSE calculation.
We present the values that meet this requirement in TABLE 
\ref{LR-tab}. With these values plus the rest of the LR 
parameters, we perform a band structure calculation.  
Results are plotted in Fig. \ref{FIG-4}a, along with a ILDOS
calculation, shown in Fig. \ref{FIG-4}b. 
It can be recognized right away that the inclusion of 
correlation gives a more localized character to the states
for both Ni $d$ orbitals and Se $p$ orbitals (A similar
calculation for NiS$_2$ is available in the SM). 
This is what is expected from previous results. Also, the
dispersion of top valence bands becomes very similar
to the hybrid bands in Fig. \ref{FIG-2}a.

The band gap dependence on $V_{NiX}$ can be linked to the 
charge transfer properties of the materials. As has been
discussed previously, without Hubbard corrections, the 
system already show an inverted valence band ordering 
similar to the case of transition metal 
oxides \cite{khomskii-book}. 
At this PBE level, if final orbital manifold occupations
are computed, it is obtained that, for Ni atoms, there exists
a considerable outer shell occupation of $\sim 9.20$. 
Formally, Ni must be in a state with $d^6$. On the other
hand, the $p$-shell of each Se (S) atom finishes with
an occupation of 4.40. Formally, the initial state of
this manifold is $p^6$. Thus we have evidence that
a charge transfer has occurred, where the $p$-shell 
occupation of each chalcogen atom has changed by
approximately 1.5.
This implies a $p$-hole formation, a well-known
phenomenon in these charge transfer 
systems \cite{Nio_PBR2020}. 
The inclusion of correlations alters occupations
slightly. But the tendency gives us clues about 
the behavior of the electronic structure. 
Thus, if we start from the LR original state
(no parameter modification), 
the occupation of the Ni $d$ shell decreases and increases
for the Se (S) $p$-shell. 
This is understandable because the rise in 
electronic localization hinders charge transfer. 
However, the system remains in the charge 
transfer regime as $p$ orbitals continue to be
higher in energy with respect to $d$ orbitals. 
Now when $V_{NiX}$ departs from the LR result to higher
values, if the state is still in the FiM state, the 
charge transfer process shows an opposite pattern of 
evolution for different atoms and spins. 
That is to say, for the chalcogen atoms, the majority
spin gets its occupations diminished while the 
minority achieves greater occupations. In the case of 
the Ni atoms, the behavior gets reversed. 
This mechanism conflates so that the spin unbalance
is utterly canceled. 
Once inside the non-magnetic phase, the augmentation of
$V_{NiX}$ has the effect of reducing the charge 
transfer. This reduction is coupled with a new 
increase in localization, which competes with 
the orbital mixing between Ni and Se (S) atoms,
The former is traduced in energy level 
repulsion. This implies that bonding $d$ orbitals 
will get deeper in energy, separating from the bands 
with a predominantly $p$-type character. Thus, the 
corresponding antibonding orbitals associated with
the lower conduction bands will 
increase their energy. Overall, the band
gap grows, controlled by the value of the intersite
interaction. In other words, the increase of $V_{NiX}$
will make energetically more costly the generation
of charge transfer excitations, which traduces 
in a greater band gap.

\section{Discussion and concluding remarks}

To put in perspective the findings obtained,
we can make a few additional remarks. First, 
although the PBE level already shows the distinctive 
indicators of the charge transfer state, such as 
the inverted band ordering, correlations must be 
included in this type of system since it is a crucial
competing mechanism for covalency and orbital
hybridization.
As shown above, it is a correlation that defines the
magnitudemof the energy band gap, which is important,
for example, to study optical responses.
Regarding the LR calculation, although this phase does
not seem to be the one that experimentally materializes 
\cite{NiSe2_exp}, we found that the calculation is useful
to get an idea of the order of magnitude of the parameters
that the systems could have. 
Also, considering the technical part, 
we have tried to amend the outcome by starting
from different insulating and metallic states, and 
ultimately the calculation consistently tends to the
featured magnetic phase. This could be linked to an
already recognized issue related to calculations on 
systems with a nearly full 
shell \cite{LR-Timrov,HP_code}.
Despite that, the minimal modification presented here
is a valuable prospect for performing future physical
response computations involving electronic correlation.
Viewing the entire array of results, if the insulating 
phase is corroborated in subsequent studies, the charge 
transfer phase will be an interesting system to analyze 
further. This is because, as has been identified in other
works, the $p$-hole formation, also known as self-doping, 
could lead to other effects when the system is perturbed 
\cite{CTI_PRL04,NiO_PRL07,khomskii-book,Nio_PBR2020}.
The interplay between the two-dimensional character 
of the materials and this $p$ orbital dominance 
could also lead to novel phenomena. 

In summary, we have put forward a detailed analysis of
the electronic structure of monolayer transition metal
dichalcogenides, NiS$_2$ and NiSe$_2$ using a variety
of first-principles methods. This allows us to ponder 
the role that electronic interaction plays at the 
mean-field level. An hybrid HSE functional was used as 
the first reference in regard that the inclusion
of correlations is built-in in its formulation.
Comparison with modern Hubbard corrected PBE
functionals allowed a further understanding of
the importance of localization and the relation
to the charge transfer effect. 
We demonstrate that nearest-neighbor intersite
parameter $V_{NiS}$ is directly linked to the 
the energy band gap of the ground state, as this 
parameters control the Ni-X hybridization and the 
charge transfer excitations.
In the future, further analysis of this problem
will focus on enhancing the inclusion of 
correlation through more advanced calculations,
such as GW calculations. Investigating how the
DFT+U+V method measures up against other
state-of-the-art approaches would be quite
intriguing. Ultimately, our goal is to highlight
the importance of including more interaction
parameters in studies like this. Doing so could
offer new insights into the behavior of
correlated systems.

\section*{Acknowledgements}
This work has been supported by the Postdoctoral Grant from 
Universidad T\'ecnica Federico Santa Mar\' ia, 
and Chilean FONDECYT Grant 1220700.  

\bibliography{main}

\end{document}


\author{\text{Sergio Bravo}$^\dagger$}
\author{P.A. Orellana}
\author{L. Rosales}
\affil[$\dagger$]{sergio.bravoc@usm.cl}
\date{}
\maketitle

\section*{Supplementary Tables}

\begin{table}[ht]
\resizebox{\columnwidth}{!}{%
\begin{tabular}{cccccc}
\hline
Material    & Lattice constant  & Ni-X distance & X-X distance & X-Ni-X angle & \multicolumn{1}{l}{X vertical distance} \\ \hline
NiSe$_2$ NM & 3.540 {\AA} & 2.390 {\AA} & 3.211 {\AA} & 95.58\degree  & 1.238 {\AA} \\
NiS$_2$  NM & 3.306 {\AA} & 2.248 {\AA} & 3.047 {\AA} & 94.68\degree  & 1.188 {\AA} \\
NiSe$_2$ LR & 3.709 {\AA} & 2.484 {\AA} & 3.304 {\AA} & 96.61\degree  & 1.258 {\AA} \\
NiS$_2$  LR & 3.549 {\AA} & 2.361 {\AA} & 3.114 {\AA} & 97.47\degree  & 1.172 {\AA}                                 
\end{tabular}%
}
\caption{Geometrical data for the NiX$_2$ materials obtained from the 
first-principles calculations. NM stands for the nonmagnetic phases presented
in the main text, with and without Hubbard corrections. LR indicates the phase 
that arises from the linear response calculation without any further 
modification. X=(S,Se) as appropriate. }
\label{TableS1}
\end{table}

\newpage
\section*{Supplementary Figures}

\begin{figure}[ht]
\centering
\includegraphics[width=1.\columnwidth,clip]{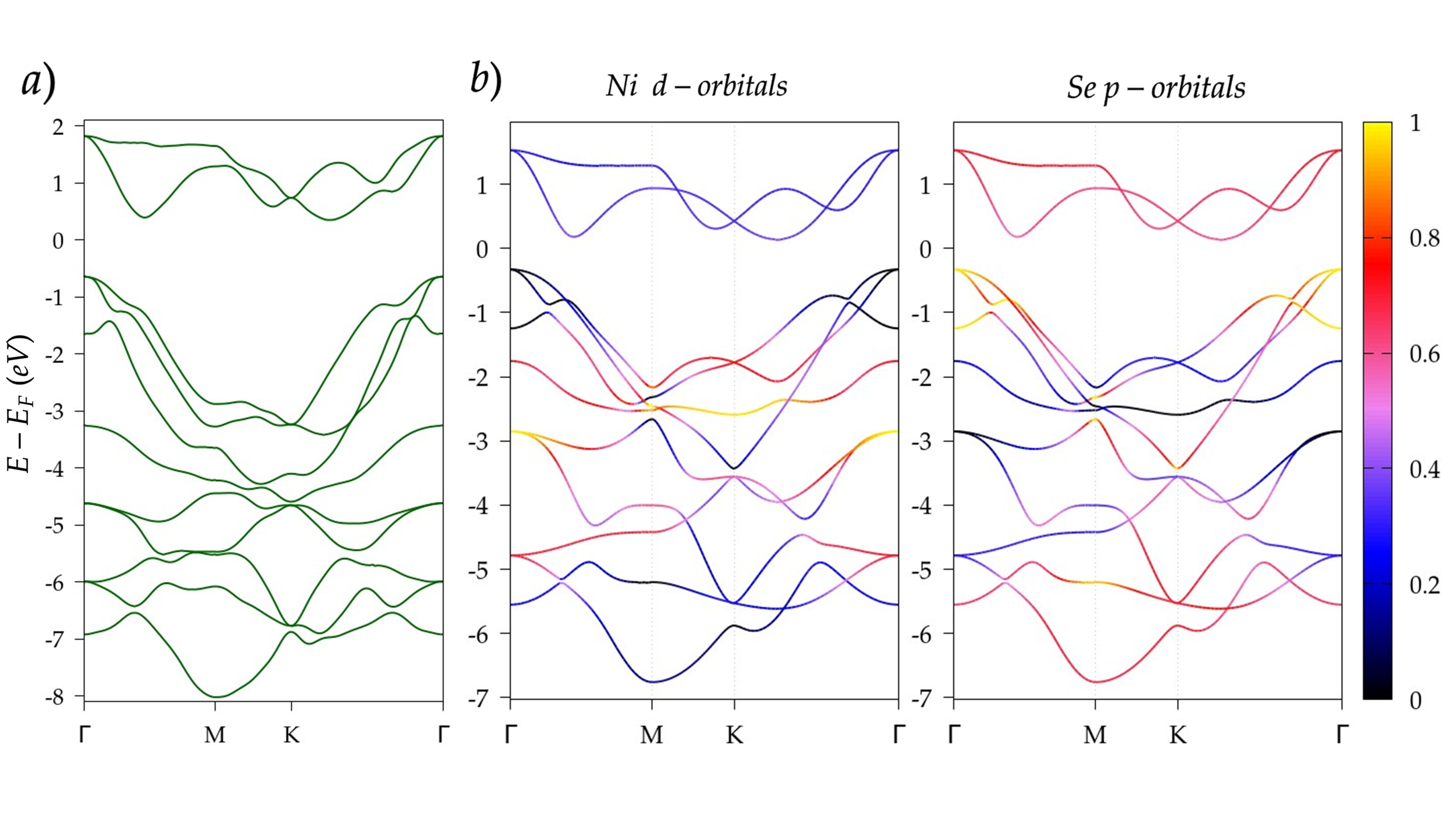}
\caption{a) HSE electronic band structure for NiS$_2$. 
b) PBE electronic band structure with orbital projections for
(left) Ni d-orbitals and (right) S p-orbitals.
The color scale at the right is in (states $\cdot$ {\AA}$^2$)/eV units.}
\label{FIG-S1}
\end{figure}

\begin{figure}[ht]
\includegraphics[width=1.\columnwidth,clip]{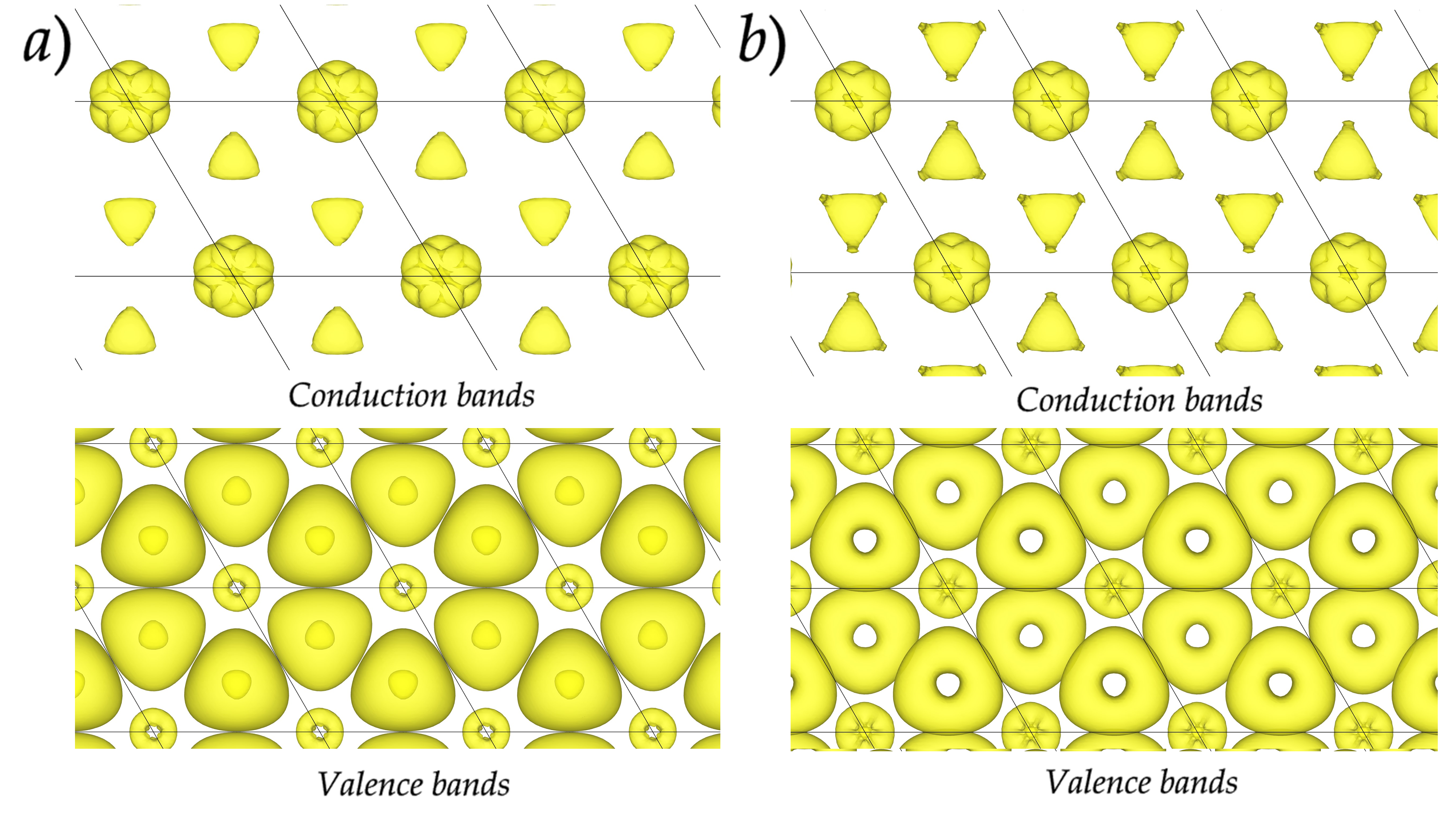}
\caption{a) Integrated local density of states for (top panel)
lowest conduction bands and (bottom panel) top of the valence 
bands with the HSE functional NiS$_2$. 
b) Integrated local density of states for (top panel)
lowest conduction bands and (bottom panel) top of the valence 
bands with the PBE functional NiS$_2$. In both cases, the 
valence band range is 0.4 eV from the top of the bands.}
\label{FIG-S2}
\end{figure}

\begin{figure}
\includegraphics[width=1.\columnwidth,clip]{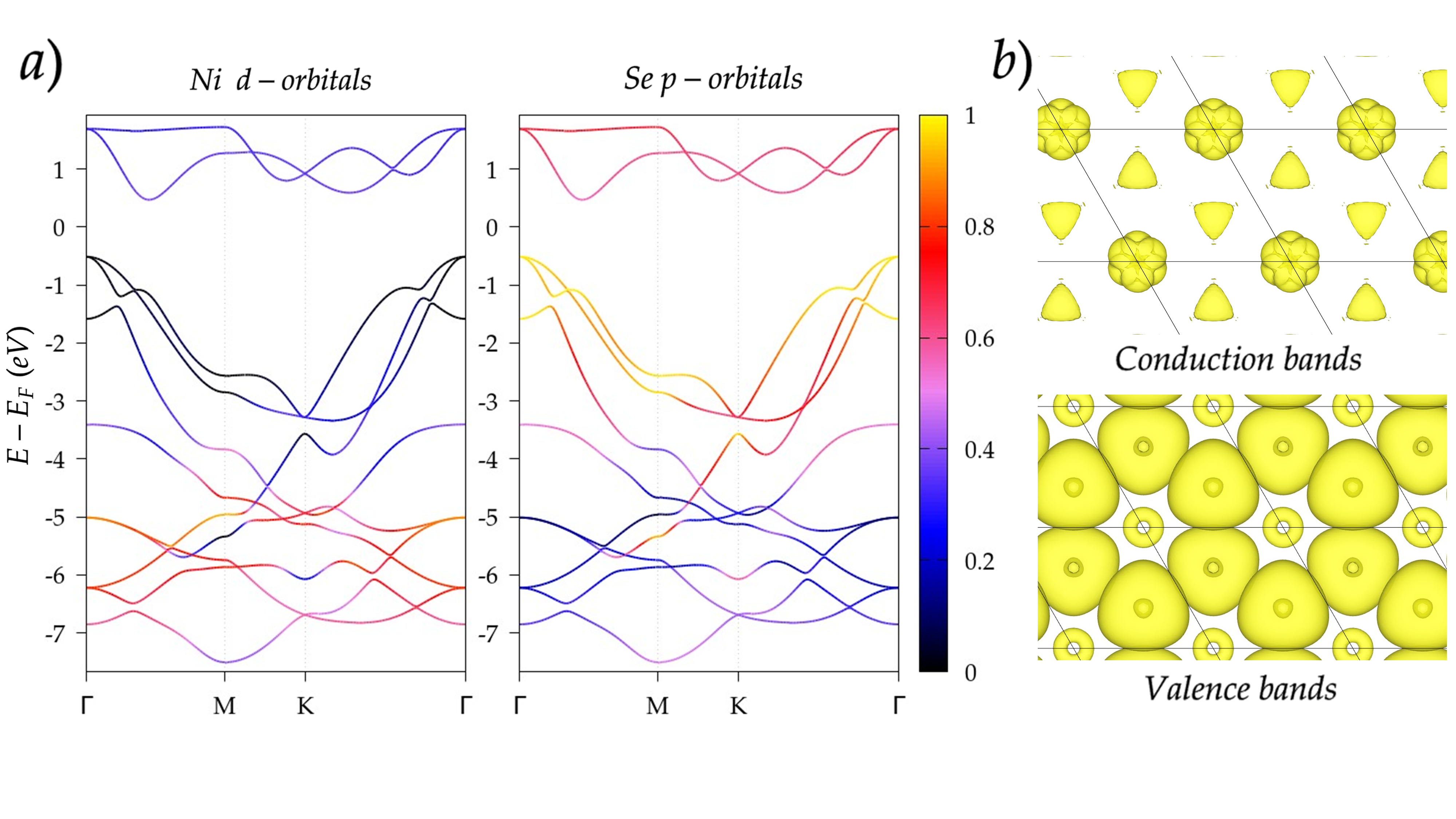}
\caption{a) Orbital-projected Electronic band structure at the
DFT+U+V level of NiS$_2$ with parameters from linear response 
with modified $V_{Ni-S}$ parameter. 
The color scale at the right is in (states $\cdot$ {\AA}$^2$)/eV units.
b) Integrated local density of states for (top panel)
lowest conduction bands and (bottom panel) top of the valence
bands using the DFT-UV functional with linear response parameters
with modified $V_{Ni-S}$.
}
\label{FIG-S3}
\end{figure}

\begin{figure}
\includegraphics[width=1.\columnwidth,clip]{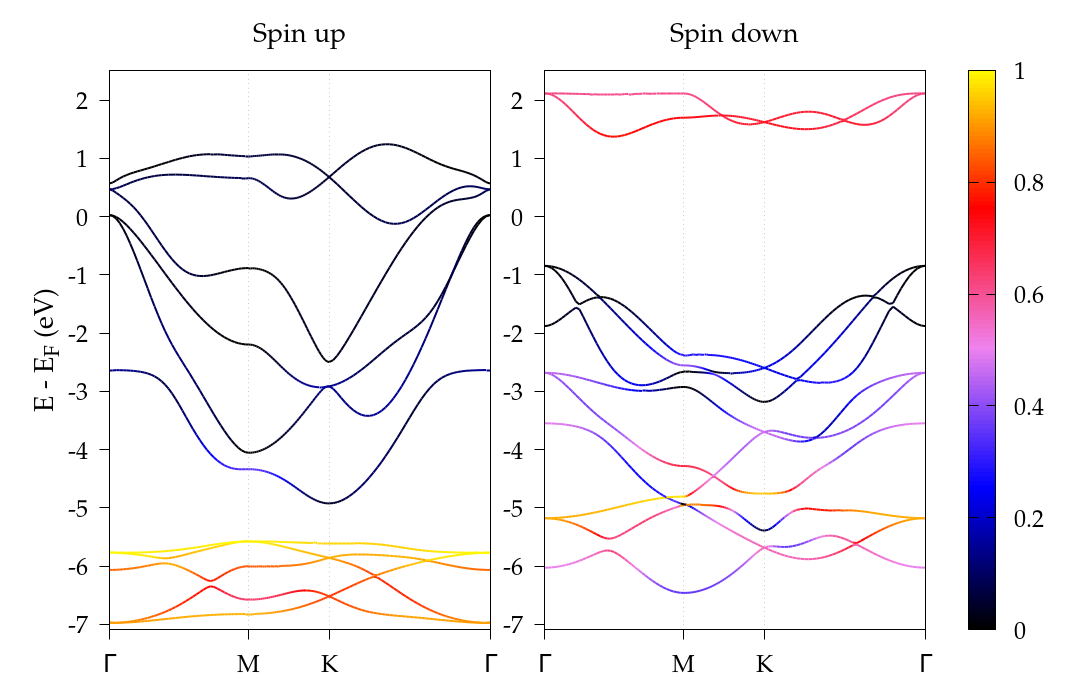}
\caption{a) Spin-polarized orbital-projected Electronic band structure
at the DFT+U+V level of Ni $d$ orbitals for NiSe$_2$ with parameters
from linear response calculation as presented in the main text. 
The color scale at the right is in (states $\cdot$ {\AA}$^2$)/eV units.
}
\label{FIG-S4}
\end{figure}

\begin{figure}
\includegraphics[width=1.\columnwidth,clip]{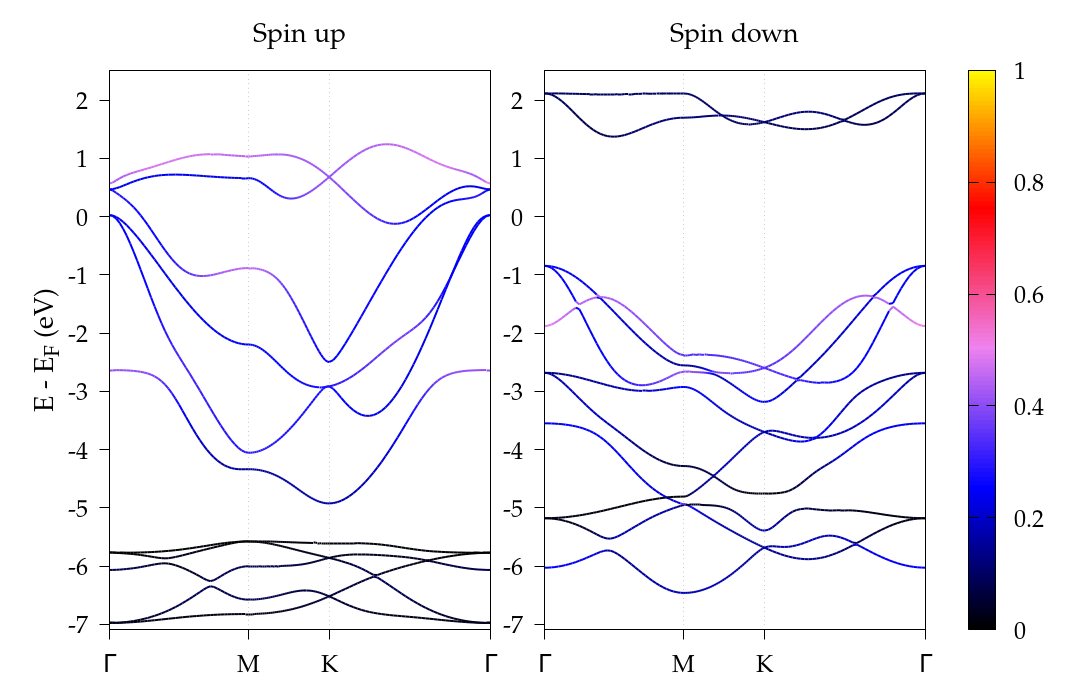}
\caption{a) Spin-polarized orbital-projected Electronic band structure
at the DFT+U+V level of Se $p$ orbitals for NiSe$_2$ with parameters
from linear response calculation as presented in the main text. 
The color scale at the right is in (states $\cdot$ {\AA}$^2$)/eV units.
}
\label{FIG-S5}
\end{figure}

\begin{figure}
\includegraphics[width=1.\columnwidth,clip]{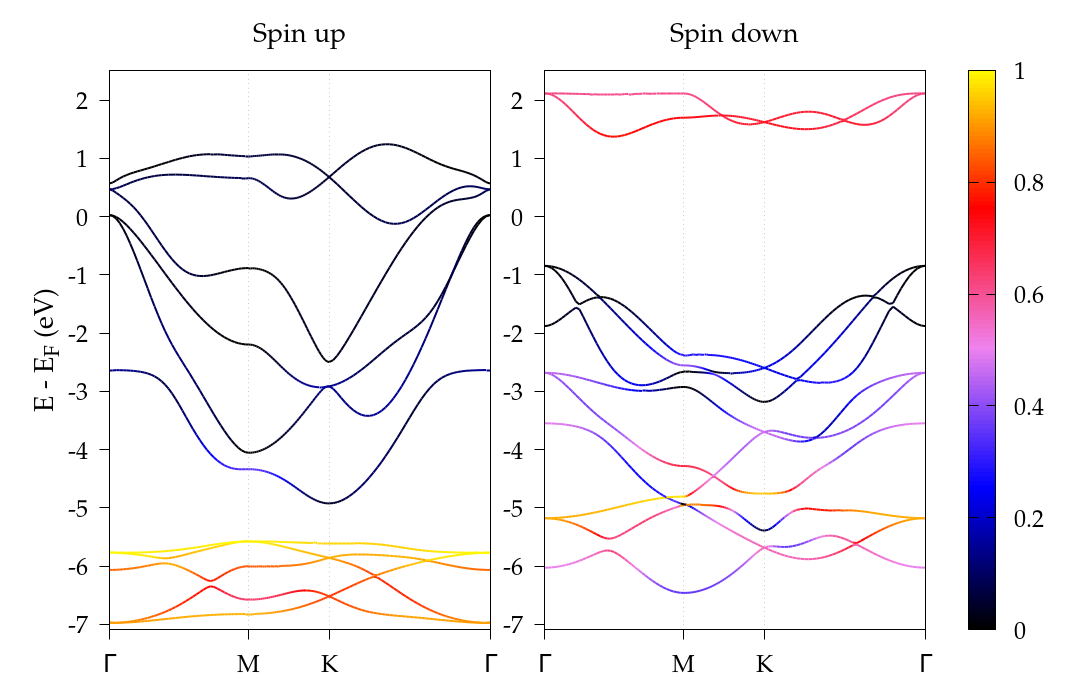}
\caption{a) Spin-polarized orbital-projected Electronic band structure
at the DFT+U+V level of Ni $d$ orbitals for NiS$_2$ with parameters
from linear response calculation as presented in the main text. 
The color scale at the right is in (states $\cdot$ {\AA}$^2$)/eV units.
}
\label{FIG-S6}
\end{figure}

\begin{figure}
\includegraphics[width=1.\columnwidth,clip]{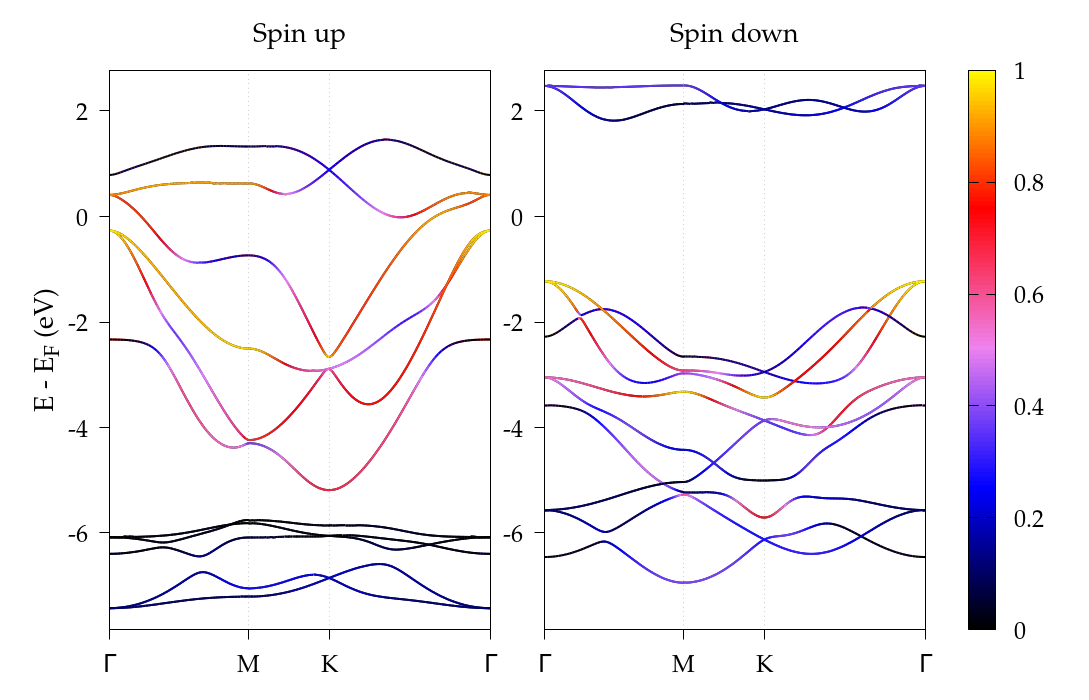}
\caption{a) Spin-polarized orbital-projected Electronic band structure
at the DFT+U+V level of S $p$ orbitals for NiS$_2$ with parameters
from linear response calculation as presented in the main text. 
The color scale at the right is in (states $\cdot$ {\AA}$^2$)/eV units.
}
\label{FIG-S7}
\end{figure}

